\documentclass[prl,twocolumn,showpacs,superscriptaddress]{revtex4}

\usepackage{graphicx}
\usepackage{amsmath}
\usepackage{amssymb}
\usepackage{amsthm}
\usepackage{mathrsfs}

\newcommand{\tr}{\operatorname{tr}}

\newtheorem{theorem}{Theorem}
\newtheorem{corollary}{Corollary}

\begin{document}
\bibliographystyle{apsrev}

\title{Entanglement for rank-$2$ mixed states}

\author{Tobias J.\ Osborne}
\email[]{tjo@maths.uq.edu.au}
\affiliation{Centre for Quantum Computer Technology and Department of Physics, 
University of
Queensland 4072, Australia}

\date{\today}

\begin{abstract}
In a recent paper, Rungta \emph{et.\ al.} 
[Phys.\ Rev.\ A, $\mathbf{64}$, 042315,
$2001$] 
introduced a measure of mixed-state
entanglement called
the \emph{$I$-concurrence} for
arbitrary pairs of qudits.  We find an exact 
formula for an entanglement measure closely related to the $I$-concurrence, the
\emph{$I$-tangle}, for all mixed states of two qudits having
no more than two nonzero eigenvalues.  We use this formula to provide a
tight upper bound for the entanglement of formation for rank-$2$ mixed 
states of a qubit
and a qudit.
\end{abstract}


\maketitle

Quantum entanglement, as typified by the singlet state
$|\Psi^-\rangle=\frac{1}{\sqrt2}(|01\rangle-|10\rangle)$, is a uniquely quantum property 
of any nonlocal superposition state.   Unlike other properties of 
quantum states, entanglement is singled out by the apparently central
role it plays in quantum information processing.  Indeed, it is becoming
clear that entanglement is a \emph{resource} which can be used to perform
quantum communication and computational tasks \cite{bennett:1996a}.  
To this end, it is of paramount importance
that a general theory of entanglement be developed in order to understand
how it can and cannot be manipulated.  The first step in this program is to
develop physically motivated \emph{entanglement measures} to quantify how
much entanglement is present in a given state.  
Many entanglement measures have been proposed, however, for mixed states,
they typically involve difficult minimisations.

The development of an efficient formula for the 
entanglement present in arbitrary 
mixed states is a crucial goal of
quantum information science.  Such a formula would 
lead to great advances in the theory of quantum communication, as well as
possibly revealing the correct role entanglement plays in quantum
computational speed-up.  To date, while there has been a great deal of
work on this subject, there is only one 
easily-computable entanglement measure and it is specific to pairs of
qubits
\cite{hill:1997a, wootters:1998a}.
The purpose of this Letter is to add to this situation by providing a 
formula for one
particular bipartite 
entanglement measure, the $I$-tangle,
for mixed states which have at most two nonzero eigenvalues.  

The structure of this Letter is as follows.  We begin by introducing 
the concurrence and the tangle for a pair of qubits. We then discuss a
generalisation of the concurrence and the tangle for pairs of qudits.  The
main result of this Letter, a formula for the $I$-tangle
of rank-$2$ mixed states, is then established.  We conclude with
a corollary of the main result, an upper bound for the entanglement
of formation of a rank-$2$ mixed state of a qubit and a qudit.  

Before we discuss the $I$-tangle, 
we introduce the \emph{concurrence}, a mixed-state entanglement measure 
for states of 
a pair of qubits $AB$ \cite{hill:1997a, wootters:1998a, wootters:2001a}.
The definition of the concurrence makes use of a specific transformation on
density operators, the \emph{spin-flip} operation, which is defined as
follows.
Consider an arbitrary mixed state $\rho$ of $AB$.
We define the spin-flip of $\rho$ to
be
\begin{equation}\label{eq:spinf}
\widetilde{\rho} \equiv \tr(\rho^\dag)I\otimes I - \rho_A^\dag\otimes I - I\otimes
\rho_B^\dag +
\rho^\dag,
\end{equation}
where $\rho_A=\tr_B(\rho)$ and $\rho_B=\tr_A(\rho)$ 
denote the reduced density operators for
subsystems $A$ and $B$ respectively. (We have included the trace and
hermitian adjoint terms so that the spin-flip operation is defined
for \emph{arbitrary} operators acting on $AB$.)  The formula for the spin-flip
is applicable
to arbitrary bipartite systems, in which case it is called the
\emph{universal state inverter} \cite{rungta:2001a}.

The spin-flip operation Eq.~(\ref{eq:spinf}) on a pair of qubits 
is an example of an antilinear operation.  
To be more precise, consider a
pure state 
$\rho=
|\psi\rangle\langle\psi|$.  The spin-flip operation, when applied to this
state, is equivalent to 
the expression $\widetilde{\rho} =
|\widetilde{\psi}\rangle\langle\widetilde{\psi}|$, where
\begin{equation}\label{eq:2qsf}
|\widetilde{\psi}\rangle = \sigma^y\otimes\sigma^y(|\psi\rangle)^*,
\end{equation}
and where $\sigma^y$ is expressed in the computational basis as 
$\left(\begin{smallmatrix}0&-i\\i&0\end{smallmatrix}\right)$,
and the complex conjugation is taken in the computational basis.
The operation in Eq.~(\ref{eq:2qsf}) is clearly an antilinear operator
\cite{endnote11}.
The definition of the spin-flip as an antilinear operator extends, via
linearity, to all mixed states,
$\widetilde{\rho} =
\overset{\rightarrow}{\theta}\rho\overset{\leftarrow}{\theta}$,
where we have added the arrows above the antilinear operator
$\theta$ representing the spin-flip 
to indicate the direction in which it acts. 
It is worth noting that 
the description of the spin-flip Eq.~(\ref{eq:spinf}) in terms of
an antilinear operator $\theta$ is specific to two qubits. 

For pure states $\rho=|\psi\rangle\langle\psi|$, the concurrence $C$ of
$|\psi\rangle$ is defined to be
$C=|\langle\psi|\widetilde{\psi}\rangle|=\sqrt{\langle\psi|\widetilde{\rho}|\psi\rangle}$.
When the state $\rho$ of the two qubits is mixed, the
concurrence $C$ is defined to be
a minimum over all pure-state
decompositions $\{p_i, |\psi_i\rangle \}$ of $\rho$:
\begin{equation}\label{eq:oldc}
C(\rho) = \min_{ \{p_i, |\psi_i\rangle \} } \sum_i p_i
|\langle\psi_i|\widetilde{\psi}_i\rangle|.
\end{equation}
It is convenient to introduce another entanglement measure closely related
to the concurrence, the \emph{tangle} $\tau$ \cite{endnote12}, 
which is also defined as a
minimisation over pure-state decompositions:
\begin{equation}\label{eq:oldt}
\tau(\rho) = \min_{ \{p_i, |\psi_i\rangle \} } \sum_i p_i
|\langle\psi_i|\widetilde{\psi}_i\rangle|^2.
\end{equation}

The squared-concurrence satisfies the inequality $C^2\le \tau$, which
follows from the convexity of 
$|\langle\psi_i|\widetilde{\psi}_i\rangle|^2=C^2(|\psi_i\rangle)$.
It turns out that the reverse inequality also holds, so  
that 
the tangle is equal to the square of the concurrence,
$\tau(\rho)=C^2(\rho)$ \cite{endnote13}.
(The reverse inequality may be established by
noting that there exists a decomposition $\{p_i,|\psi_i\rangle\}$ achieving the
minimum in Eq.~(\ref{eq:oldc}) which has the property that
$C(|\psi_i\rangle)=C(|\psi_j\rangle)$ \cite{wootters:1998a}.  
The inequality follows from
substituting this decomposition into the expressions for $\tau$ and $C^2$.)
A simple formula for the concurrence of two qubits is known
\cite{wootters:1998a},
\begin{equation}\label{eq:ccform}
C(\rho) = \max[0,\lambda_1-\lambda_2-\lambda_3-\lambda_4],
\end{equation}
where the $\lambda_i$ are the square-roots of the 
singular values, in decreasing order, of the
matrix $\rho\widetilde{\rho}$.
 
For a pair of qudits $AB$ we use a variant of the
$I$-concurrence of 
Rungta \emph{et.\ al.} \cite{rungta:2001a} to measure the entanglement for 
mixed states of $A$
and $B$.  The $I$-concurrence is defined via Eq.~(\ref{eq:spinf}) 
\cite{rungta:2001a},
\begin{equation}
C(\rho) = \min_{ \{p_i, |\psi_i\rangle \} } \sum_i p_i
\sqrt{\langle\psi_i|\widetilde{\rho_i}|\psi_i\rangle},
\end{equation}
where $\rho_i=|\psi_i\rangle\langle\psi_i|$.
The entanglement measure we use is a generalisation of the tangle, 
the \emph{$I$-tangle},
defined by
\begin{equation}\label{eq:itanglemin}
\tau(\rho) = \min_{ \{p_i, |\psi_i\rangle \} } \sum_i p_i
\langle\psi_i|\widetilde{\rho_i}|\psi_i\rangle.
\end{equation} 
The $I$-concurrence and the $I$-tangle are good mixed-state
entanglement measures because they satisfy the standard properties usually regarded
as essential for a good
entanglement measure (see, for example, \cite{vidal:2000a, vedral:1998a}).
The inequality $C^2\le \tau$ may be established, by convexity,
as for two qubits. 
Because the equal-entanglement decomposition only exists for pairs of qubits, 
the $I$-tangle is not, in general, equal to the square of the
$I$-concurrence.  Based on the results of this paper, and the calculations
of the $I$-concurrence for isotropic states  
\cite{endnote14} we feel that the $I$-tangle, 
as defined by a minimisation, is the proper generalisation of the
tangle Eq.~(\ref{eq:oldt}).   

The universal state inverter Eq.~(\ref{eq:spinf}) may be expressed in
terms of another formula which will be most useful in the following.
Before we write down this formula, however, we need to
introduce some definitions.  Let $|i\rangle_A$ and $|j\rangle_B$ denote the
computational basis states for subsystems $A$ and $B$, with dimensions 
$d_A$ and $d_B$, respectively.  
For an
arbitrary pair
$\{|i\rangle_A,|i'\rangle_A\}$, $\{|j\rangle_B,|j'\rangle_B\}$ of
the computational basis states of $AB$ we set up the
projectors $P^{(ii')}_A = |i\rangle_A\langle i| + |i'\rangle_A\langle i'|$,
$P^{(jj')}_B = |j\rangle_B\langle j| + |j'\rangle_B\langle j'|$, and 
$Q_\alpha = P^{(ii')}_A\otimes P^{(jj')}_B$, where $\alpha = (i,i',j,j')$.
Consider the object $\rho_\alpha = Q_\alpha \rho Q_\alpha$.  The operator
$\rho_\alpha$ is a positive operator supported on a $2\times 2$
subspace of the
Hilbert space of $AB$ spanned by $\{ |ij\rangle,
|i'j\rangle, |ij'\rangle, |i'j'\rangle\}$.  
In this way we can think of $\rho_\alpha$ as a  
subnormalised state of two qubits.
The two-qubit spin flip, when applied to $\rho_\alpha$, gives 
\begin{equation}
\widetilde{\rho}_\alpha = \overset{\rightarrow}{\theta}_\alpha \rho \overset{\leftarrow}{\theta}_\alpha =
\sigma^y\otimes\sigma^y (Q_\alpha
\rho Q_\alpha)^* \sigma^y\otimes\sigma^y,
\end{equation}
where $\theta_\alpha= \theta
Q_\alpha$ is the antilinear operator 
representing the spin-flip
operation on the $2\times 2$ subspace, and $\sigma^y$ is naturally defined
on the two-dimensional subspaces of $A$ and $B$ respectively.
Using these definitions we can write an alternative formula for the
universal state inverter,
\begin{equation}\label{eq:tasum}
\widetilde{\rho}=\sum_{\alpha} \overset{\rightarrow}{\theta}_\alpha \rho \overset{\leftarrow}{\theta}_\alpha,
\end{equation}
where the sum over $\alpha$ runs over all of the
$(d_A(d_A-1)/2)(d_B(d_B-1)/2)$ possible choices of pairs of computational
basis states.  (The
reader may verify that Eq.~(\ref{eq:tasum}) 
follows from the expression of the universal
state inverter as a tensor product of two superoperators of the form 
$\mathcal{P}\circ\mathcal{T}$.  See \cite{rungta:2001a} for further
details.)

It is convenient, at this point, to introduce two quantities which
will simplify the statement of our main result. 
Let $\rho$ be a density operator for a pair of qudits having no more
than two non-zero eigenvalues.  We may write $\rho$ in terms of its
eigenvectors,
\begin{equation}
\rho = p|v_1\rangle\langle v_1| + (1-p)|v_2\rangle\langle v_2|.
\end{equation}
Using these eigenvectors we construct the tensor
\begin{equation}
T_{ijkl} = \mbox{tr}(\gamma_{ij}\widetilde{\gamma}_{kl}),
\end{equation}
where $\gamma_{ij}=|v_i\rangle\langle v_j|$.  We also construct the 
real symmetric $3\times3$ matrix $M_{ij}$ whose independent 
entries are given by
\begin{equation}\label{eq:mmatrix}
\begin{split}
M_{11} &= \frac14T_{1221} + \frac12T_{1122} + \frac14T_{2112}, \\
M_{12} &= \frac i4T_{1221} - \frac i4T_{2112}, \\
M_{13} &= \frac14T_{1121} - \frac14T_{2122} + \frac14T_{1112} -
\frac14T_{1222}, \\
M_{22} &= -\frac14T_{1221} + \frac12T_{1122} -\frac14T_{2112}, \\
M_{23} &= \frac i4T_{1121} - \frac i4T_{1112} + \frac i4T_{2122} -
\frac i4T_{1222}, \\
M_{33} &= \frac14T_{1111} - \frac12T_{1122} + \frac14T_{2222}.
\end{split}
\end{equation}
(The entries of $M$ will be shown to be real in the following.)

We now have all the necessary ingredients required for the statement of our
main result.
\begin{theorem}
Let $\rho$ be any density operator for a pair $AB$ of qudits, of dimensions
$d_A$ and $d_B$, respectively, having no more than
two nonzero eigenvalues.  The $I$-tangle $\tau$ between $A$ and $B$ is
given by the expression 
\begin{equation}\label{eq:itangle}
\tau(\rho) = \tr(\rho\widetilde{\rho}) +
2\lambda_{\min}(1-\tr(\rho^2)),
\end{equation}
where $\lambda_{\min}$ is the smallest eigenvalue of the matrix $M$
defined by Eq.~(\ref{eq:mmatrix}).
\end{theorem}
The formula Eq.~(\ref{eq:itangle}) for the $I$-tangle is easy to compute
for all rank-$2$ mixed states of a pair of qudits. 
We have verified this formula by performing the minimisation in
Eq.~(\ref{eq:itanglemin}) numerically for a large number of random states,
and compared the results with the
formula Eq.~(\ref{eq:itangle}).  We found agreement to 
within $10^{-16}$.  
\begin{proof}
The method we use to prove this theorem is similar to that employed by
Hill and Wootters \cite{hill:1997a}.

Consider an arbitrary pure state $|\psi\rangle$ which can be written as a
linear combination of the two eigenvectors of $\rho$, $|\psi\rangle =
c_1|v_1\rangle + c_2|v_2\rangle$.  The $I$-tangle of $|\psi\rangle$ is
given by the expression
\begin{equation}
\tau(\psi) = \langle\psi|\widetilde{\sigma}|\psi\rangle = \sum_\alpha
\langle\psi|\overset{\rightarrow}{\theta}_\alpha|\psi\rangle\langle\psi|\overset{\leftarrow}{\theta}_\alpha|\psi\rangle,
\end{equation}  
where $\sigma = |\psi\rangle\langle\psi|$, and we have used Eq.~(\ref{eq:tasum})
to rewrite the spin flip in terms of the antilinear operators
$\theta_\alpha$.  
Each of the terms in the sum over $\alpha$ may be written as a trace,
\begin{equation}\label{eq:csq1}
\tau(\psi) = \sum_\alpha\tr\left(\omega^* \zeta^{\alpha} \omega
{\zeta^{\alpha}}^*\right),
\end{equation}
where $\omega_{ij}=c_ic_j^*$ 
is the density matrix of $|\psi\rangle$ expressed in the
$\{|v_1\rangle,|v_2\rangle\}$ basis, and $\zeta^{\alpha}_{ij} = \langle
v_i|\overset{\rightarrow}{\theta}_\alpha|v_j\rangle$.

The function on the RHS of Eq.~(\ref{eq:csq1}) can be extended via linearity
to a function $f$ of all $2\times2$ density matrices $\omega$ expressed in
terms of the $\{|v_1\rangle,|v_2\rangle\}$ basis, i.e., $f(\omega) =
\sum_\alpha\tr\left(\omega^* \zeta^{\alpha} \omega
{\zeta^{\alpha}}^*\right)$.  The function $f$ has the property that it is
equal to the $I$-tangle for all pure states $\psi$, $f(\psi)=\tau(\psi)$.

Any $2\times2$ density operator $\omega$ may be expressed in terms of the Pauli
matrices via the operator expansion,
$\omega=\frac12(I+\mathbf{r}\cdot\boldsymbol{\sigma})$, where
$r_i=\tr(\omega\sigma^i)$.  Substituting this expansion into the expression
for $f$ gives
the quadratic form
\begin{equation}\label{eq:ffun}
f(\omega) = \frac14\tr(\Upsilon) +
\sum_j r_j L_j + \sum_{j,k}r_jr_k M_{jk},
\end{equation}
where $\Upsilon = \sum_{\alpha}{\zeta^\alpha}^*\zeta^\alpha$,
\begin{equation}\label{eq:lj}
L_j = \tr(\sigma^j\Upsilon),
\end{equation}
and
\begin{equation}\label{eq:mmatrix2}
M_{jk} = \sum_{\alpha}
\tr({\sigma^j}^*\zeta^\alpha\sigma^k{\zeta^\alpha}^*).
\end{equation}
Each of the terms in the sum over $\alpha$ in Eq.~(\ref{eq:mmatrix2}) is a
real symmetric matrix, so that $M$ is a real symmetric matrix.  It may be
straightforwardly verified that the entries
of $M$ are given by Eq.~(\ref{eq:mmatrix}).

For the rank-$2$ density operator $\rho$, 
the state-space of the system $AB$ can be considered to be the space of all
convex combinations of superpositions of $|v_1\rangle$ and $|v_2\rangle$.
If a particular state $|\psi\rangle$ of $AB$ is pure, its corresponding
$2\times2$ density operator in the $\{|v_1\rangle$, $|v_2\rangle\}$ basis,
$\omega=\frac12(I+\mathbf{r}\cdot\boldsymbol{\sigma})$, satisfies the
condition $|\mathbf{r}|^2=1$. 
In this way, we can think of the entire state-space as the Bloch sphere where 
the poles are the eigenvectors $|v_1\rangle$ and $|v_2\rangle$.  A 
particular decomposition of $\rho$ may be viewed as the weighted
sum of points on the surface of the Bloch sphere, where $\rho$ lies at the
centre of mass of the weighted sum.
The function $f$ is defined on the entire state-space 
$|\mathbf{r}|^2 \le 1$.  
 
When the bipartite system $AB$ is a pair of qubits, there is only one term
in the sum Eq.~(\ref{eq:tasum}), and $f$ 
reduces to the quadratic form  
that Hill and Wootters \cite{hill:1997a} study.  In this case, the
eigenvalues of $M$ are given by $\pm\frac12|\det\zeta|$ and
$\frac14\tr(\zeta^*\zeta)$, which means that $f$ is convex along two directions and concave along a third.  In general, the matrix $M$ will have three
positive eigenvalues, so that $f$ is typically
convex.

For the purposes of this proof it is essential that a quadratic form $g$ be
constructed which agrees with $f$ on pure states $\psi$ which has the
additional 
property that it is convex along two directions and linear along a third.
A function $g$ which has these properties may be constructed from $f$ 
as follows,
\begin{equation}
g(\omega) \equiv f(\omega) - \lambda_{\text{min}}(|\mathbf{r}|^2-1),
\end{equation}
where $\lambda_{\min}$ is the smallest eigenvalue of the matrix $M$.
This function is a quadratic form, 
\begin{equation}
g(\omega) = K + \sum_j r_j L_j + \sum_{j,k}r_jr_kN_{jk},
\end{equation}
where $N=M-\lambda_{\text{min}}I$, and $K = \frac14\tr(\Upsilon) +
\lambda_{\text{min}}$.  The matrix $N$ which defines the quadratic form $g$
has two positive eigenvalues and one zero eigenvalue so that $g$ is convex
along two directions and linear along the third.  
The quadratic form $g$ has the additional property that it is 
equal to $f$ for pure states $\psi$, ($|\mathbf{r}|^2=1$).

At this point we recall a theorem due to Uhlmann \cite{uhlmann:2000a}, which
concerns functions of density matrices 
expressed as minimisations over all pure-state
decompositions.  
\begin{theorem}\label{thm:uhlmann}
Let $G$ be a positive real-valued function defined on pure states.  The
function $\mathscr{G}$, defined for all mixed states $\rho$, given by
\begin{equation}
\mathscr{G}(\rho) = \min_{\{p_i,|\psi_i\rangle\}}\sum_i p_i G(\psi_i),
\end{equation}
where the minimisation runs over all pure-state decompositions of $\rho$,
$\{p_i,|\psi\rangle\}$,
is the \emph{largest} convex function which agrees with $G$ on pure states
$\rho=\psi$.
\end{theorem} 
The $I$-tangle is expressed as a minimisation over all pure-state
decompositions of a density operator, so if we could find the largest
convex function which agrees with $f=\tau$ on pure states $\psi$ it is
guaranteed to be equal to the $I$-tangle for all mixed states.  We claim
that $g$ is precisely this function.  To prove this statement we proceed
via contradiction.  Assume that there is a convex function $g'$ which
agrees with $f$ on pure states but which is larger than $g$.  We can write
$g'$ in terms of $g$ and a `correction' $t$
\begin{equation}
g'(\omega) = g(\omega) + t(\omega).
\end{equation}
In order that $g'$ satisfy the correct boundary
conditions $t(\omega)$ must be zero on the set of all pure states (the surface
of the Bloch sphere). 
Consider the coordinate system $x_i$ defined by the three
eigenvectors of $N$ where, without loss of generality, we choose the $x_3$ 
coordinate to correspond with the
direction along which $g$ grows linearly.  
The condition that $g'$ be convex is equivalent to
requiring that the Hessian matrix 
\begin{equation}
\mathbf{H} = \frac{\partial^2 g'}{\partial x_i\partial x_j}
\end{equation}
is positive semidefinite \cite{rockafellar:1970a}.  In order that
$\mathbf{H}$ be positive semidefinite, it is necessary (although not sufficient)
that the entry $\mathbf{H}_{33}$ 
satisfies the inequality $\mathbf{H}_{33}\ge0$ 
\cite{horn:1990a}.  Because $g$ grows linearly along $x_3$, the condition
that $\mathbf{H}_{33}\ge0$ becomes $\frac{\partial^2 t}{\partial x_3^2} \ge
0$, which means that $t$ is convex as a function of $x_3$.  
The boundary
conditions, $t(\psi)=0$, therefore imply that $t$ is everywhere negative.  Hence
$g'\le g$, and we have a contradiction.  This implies 
that $g$ is the largest convex function which takes the values
$\tau(\psi)$ on the set of all pure states.  
Therefore, theorem~\ref{thm:uhlmann} shows that $g$ is 
equal to the $I$-tangle.  

The expression for the $I$-tangle $\tau$ may be simplified by noting
that, for rank-$2$ $\rho$, 
$1-|\mathbf{r}|^2=2(1-\tr(\rho^2))$.  Note, also,  
that $f(\rho) = \tr(\rho\widetilde{\rho})$.  Hence we can write
$\tau(\rho)=\tr(\rho\widetilde{\rho})+2\lambda_{\text{min}}(1-\tr(\rho^2))$.
\end{proof}


The decomposition which achieves the minimum for the $I$-tangle
Eq.~(\ref{eq:itangle}) consists of two terms.  In contradistinction 
to the case of two
qubits, the minimising decomposition will, in general, 
consist of terms with differing values of $\tau$.  This is because the
surfaces of constant $g$ 
will typically be curved, 
so that the
trick of Hill and Wootters cannot be applied 
(see \cite{hill:1997a} for the
construction of the minimising decomposition when the surfaces of constant
$g$ are elliptic cylinders).  
The construction of the minimising decomposition follows from observing
that the function $g$ has the
property that it grows linearly in one direction.  Consider the line
parallel to the eigenvector of $N$, whose associated eigenvalue is zero,
which passes through the density operator $\rho$.  The
density operator $\rho$ may be written as a convex sum of the two pure states
$|\phi_1\rangle$ and $|\phi_2\rangle$
which lie at either end of the line, $\rho =
q_1|\phi_1\rangle\langle\phi_1|+q_2|\phi_2\rangle\langle\phi_2|$.  
Because the $I$-tangle is convex, we obtain the inequality
\begin{equation}\label{eq:optdec}
\tau(\rho) \le q_1 \tau(\phi_1)  + q_2\tau(\phi_2).
\end{equation}
However, $\tau=g$ varies linearly in this direction, so that the inequality
in Eq.~(\ref{eq:optdec}) is actually an equality.  

When one of the subsystems of the bipartite system $AB$ is a qubit it is
possible to obtain a relation between the $I$-tangle $\tau$ and the
entanglement of formation $\mathscr{F}$.  For pure states $\psi$ of $AB$
the entanglement of formation is given in terms of the $I$-tangle via
\begin{equation}
\mathscr{F}(\psi) = \mathscr{E}(\tau(\psi)),
\end{equation}
where $\mathscr{E}(x) = H\left(\frac12 + \frac12\sqrt{1-x}\right)$, and
$H$ is the binary entropy function $H(x)=-x\log x - (1-x)\log(1-x)$, where
the logarithm is taken to base $2$.  The
function $\mathscr{E}$ is concave and monotone increasing.  If we consider
the minimising decomposition $\{q_i,|\phi_i\rangle\}$ we constructed
in the previous paragraph, we obtain the chain of inequalities
\begin{equation}
\mathscr{F}(\rho) \le q_1 \mathscr{E}(\tau(\phi_1)) +
q_2\mathscr{E}(\tau(\phi_2)) \le \mathscr{E}(\tau(\rho)),
\end{equation}
where the first inequality follows from the definition of the entanglement
of formation, and the second from the fact 
$\mathscr{E}(g)$ is concave along the line passing through the
pure states $|\phi_i\rangle$.  
This statement 
is the content of
the following corollary:   
\begin{corollary}
For rank-$2$ mixed states $\rho$ of a qubit $A$ and a qudit $B$,
the entanglement of formation $\mathscr{F}$ of $\rho$ satisfies the
inequality
\begin{equation}\label{eq:eofin}
\mathscr{F}(\rho) \le H\left(\frac12 + \frac12\sqrt{1-\tau(\rho)}\right).
\end{equation}
\end{corollary}
Numerical experiments indicate that 
the expressions on the LHS and RHS of Eq.~(\ref{eq:eofin})  
usually differ only by about $10^{-4}$, so that the inequality is typically
very close to an equality; it is not, however, an equality.

We have found a formula for the $I$-tangle for all states of a pair of
qudits having no more than two nonzero eigenvalues.  We have also found an
upper bound for the entanglement of formation between rank-$2$ mixed states
of a qubit and a qudit.  The method we employed to construct these formulae
relied crucially on theorem~\ref{thm:uhlmann}.  It is tempting to
conjecture
that the construction of entanglement measures might be simplified by 
searching for large convex functions which satisfy certain boundary
conditions --- perhaps as solutions to certain partial differential 
equations.  There are some difficulties with this suggestion, however.  We
must include in this search space functions which are not analytic at one
or more points (the concurrence is not analytic at all points).  Nonetheless,
perhaps there exists a systematic way to construct good analytic 
approximations to entanglement measures expressed as minimisations over
pure state decompositions. 

\begin{acknowledgments}
I would especially like to thank Carl Caves, Michael Nielsen, 
and Bill Wootters for 
their help and for 
inspirational discussions which led to this work.  I would also like to thank
Carl Caves for pointing out the substantially simpler 
proof of Eq.~(\ref{eq:tasum}).
Thanks also to Michael Bremner, Chris Dawson, Jennifer Dodd, Alexei
Gilchrist, Duncan Mortimer, and Armin Uhlmann 
for helpful and stimulating discussions.  
This work has been funded, in part, by an Australian Postgraduate Award.
\end{acknowledgments}


\begin{thebibliography}{10}
\bibitem{endnote11}{An antilinear operator $\theta $ is an operator which satisfies $\theta (c_1|\phi _1\delimiter "526930B +c_2|\phi _2\delimiter "526930B ) = c_1^*\theta |\phi _1\delimiter "526930B +c_2^*\theta |\phi _2\delimiter "526930B $.}
\bibitem{endnote12}{It should be noted that our definition of the tangle differs from the definition used in the literature, where the tangle is simply defined to be \protect \emph  {equal} to the squared-concurrence, $\tau =C^2$. As we show, our definition is equivalent to the standard one for two qubits. We use this alternative definition because it naturally generalises to a quantity which is different to the squared-concurrence for pairs of qudits.}
\bibitem{endnote13}{The equality of the tangle and the squared-concurrence was pointed out by Michael Nielsen.}
\end{thebibliography}
\end{document}